# Effects of primitive photosynthesis on Earth's early climate system

Kazumi Ozaki[1,2,3*], Eiichi Tajika[4], Peng K. Hong[5], Yusuke Nakagawa[6], and Christopher T. Reinhard[1,2]

[1]School of Earth and Atmospheric Sciences, Georgia Institute of Technology, Atlanta, GA 30332, USA

[2]NASA Astrobiology Institute, Alternative Earths Team. Mountain View, CA 94043, USA

[3]NASA Postdoctoral Program, Universities Space Research Association, Columbia, MD 21046, USA

[4]Department of Earth and Planetary Science, Graduate School of Science, The University of Tokyo, Bunkyo-ku, Tokyo 113-0033, Japan

[5]Department of Systems Innovation, School of Engineering, The University of Tokyo, Bunkyo-ku, Tokyo 113-8656, Japan

[6]Analysis Engineering Department, Hitachi Power Solutions CO.,Ltd., Hitachi-Shi, Ibaraki 317-0073, Japan

[*]Corresponding author. E-mail: kazumi.ozaki@eas.gatech.edu

**Abstract**: **The evolution of different forms of photosynthetic life has profoundly altered the activity level of the biosphere, radically reshaping the composition of Earth's oceans and atmosphere over time. However, the mechanistic impacts of a primitive photosynthetic biosphere on Earth's early atmospheric chemistry and climate are poorly understood. Here, we use a global redox balance model to explore the biogeochemical and climatological effects of different forms of primitive photosynthesis. We find that a hybrid ecosystem of $H_2$-based and $Fe^{2+}$-based anoxygenic photoautotrophs—organisms that perform photosynthesis without producing oxygen—gives rise to a strong nonlinear amplification of Earth's methane ($CH_4$) cycle, and would thus have represented a critical component of Earth's early climate system before the advent of oxygenic photosynthesis. Using a Monte Carlo approach, we find that a photosynthetic hybrid biosphere widens the range of geochemical conditions that allow for warm climate states well beyond either of these metabolisms acting in isolation. Our results imply that the Earth's early climate was governed by a novel and poorly explored set of regulatory feedbacks linking the anoxic biosphere and the coupled H, C, and Fe cycles, with important ramifications for the sustained habitability of reducing Earth-like planets hosting primitive photosynthetic life.**

Standard stellar evolution models predict that the Sun during the Archaean eon (4.0–2.5 Gyr ago) was ~20–25% dimmer than it is today[1-3]. Despite receiving less energy from the Sun, geologic records indicate that the Earth's climate at that time was generally warm, perhaps even warmer than that of the modern Earth. This apparent paradox, the so-called 'Faint Young Sun Paradox (FYSP)'[4], has long been a topic of debate, because its resolution bears important ramifications for the basic factors structuring climate regulation and the long-term habitability of Earth and Earth-like exoplanets. Most common solutions to the FYSP invoke a stronger greenhouse effect, and rely on elevated atmospheric levels of greenhouse gasses ($NH_3$[4-6], $CO_2$[7-10], $CH_4$[11-14], $C_2H_6$[15], $N_2O$[16, 17], and/or $COS$[18]) and/or the radiative impact of $N_2$ and $H_2$ through pressure-/collision-induced absorption[4, 19-21]. However, the abundance of these gases in the atmosphere is controlled by photochemical, biological and geological processes, and it remains unclear whether any of these gases could have been sustained at abundances high enough to resolve the FYSP[22, 23]. A full understanding of this question requires a quantitative framework that incorporates the many interactions at play between the biosphere, surface geochemical cycles, and climate.

In particular, the activity level of the biosphere is a critical factor controlling the composition of Earth's atmosphere and thus the atmospheric greenhouse effect. On the modern Earth ecosystems are driven principally by the activity of oxygenic photosynthesis, which currently maintains a large net primary production flux of ~105 Gt C per year[24, 25] along with a stoichiometric release of molecular oxygen to Earth surface environments. In contrast, prior to the advent of oxygenic photosynthesis Earth's biosphere would have been driven largely by anoxygenic photosynthetic metabolisms in the sunlit surface ocean[26]. In strong contrast to oxygenic phosynthesizers, which exploit the ubiquitous $H_2O$ molecule as their electron donor, the activity of anoxygenic phototrophs seems likely to be limited by the availability of reduced substrates (including $H_2$, $Fe^{2+}$, and $H_2S$)[26-30]. Given probable limits on source strength of the single limiting substrates[28, 29], overall rates of primary production in the primitive photosynthetic biosphere would have been strongly attenuated relative to those achievable by an oxygenic biosphere[26-29], with potentially important consequences for the fluxes of many biogenic and climatologically important gases to the atmosphere.

However, previous work attempting to reconstruct Earth's primitive biosphere has typically considered individual ecosystems in isolation, and has not explored the possible large-scale synergistic effects between multiple photosynthetic and fermentative metabolisms[28, 29]. For example, because $Fe^{2+}$-using anoxygenic photoautotrophs (photoferrotrophs) utilize $H_2O$ (not $H_2$) as their hydrogen source, once the organic matter produced by photoferrotrophs is converted to $CH_4$ (and $CO_2$) by fermenters and methanogens it can be converted in the atmosphere to $H_2$ via photolysis, thus promoting the activity of $H_2$-using photosynthesis and subsequent biogenic $CH_4$ production (see Methods). The coupled effect of these primitive photosynthetic metabolisms would thus be a critical component of the surface $CH_4$ cycle, and would contribute to the regulation of the climate system through modulation of atmospheric $CH_4$ levels.

In order to quantitatively evaluate the biogeochemical and climatological effects of different primitive photosynthetic biospheres, we employ a novel coupled atmosphere-ecosystem model. The photochemical model is a one-dimensional model[12] that includes 73 chemical species involved in 359 chemical reactions (see Methods). The photochemical model is then coupled with an ocean ecosystem module, which includes a series of biogeochemical processes such as photosynthesis, methanogenesis, and the burial of organic matter and ferric oxides in sediments (Fig. 1). We focus here on two idealized scenarios with different configurations of the primitive photosynthetic biosphere (see also Supplementary Discussion). In Case 1, $H_2$-using anoxygenic phototrophs and CO-consuming acetogens are considered as producers, while fermentors and acetotrophic methanogens are included as decomposers. Case 2 represents a 'hybrid' photosynthetic biosphere, and includes $Fe^{2+}$-using anoxygenic phototrophs and dissimilatory Fe(III) reducing bacteria in addition to the same four organisms as in Case 1.

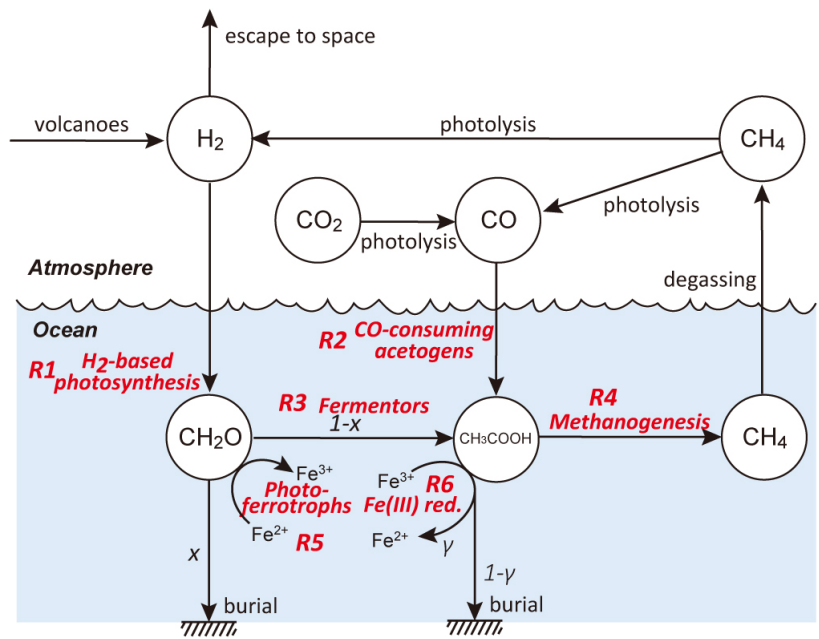

**Fig. 1. Schematic diagram of the primitive biosphere considered in this study.** The primary source of reducing power (in terms of $H_2$) is outgassing from the solid Earth. A fraction of organic matter produced by $H_2$-based (R1) and $Fe^{2+}$-based (R5) anoxygenic photosynthesis, $x$, is buried in sediments and the remaining fraction is converted to $CH_4$ (and $CO_2$) by fermentors (R3), Fe-reducers (R6) and methanogens (R4). A certain fraction of iron-oxides produced by photoferrotrophs, $\gamma$, is reduced to $Fe^{2+}$ via Fe-reduction before burial. $CH_4$ generated by methanogenesis liberates to the atmosphere and is converted to $H_2$ via photolysis reaction in the atmosphere. See Supplementary Methods for further details.

Surface temperature is evaluated offline based on the results of a 1-D radiative-convective climate model[15] in which the solar luminosity is set to 80% of its present value (e.g., roughly mid-Archaean time), and includes the radiative effects of $C_2H_6$. Our model is thus meant to represent a generalized mid-Archaean Earth system without oxygenic photosynthesis, or a potential Earth-like exoplanet around a young G-type star with a primitive photosynthetic biosphere. We note, however, that there is isotopic evidence[31, 32] suggesting the presence of oxygenic photosynthesis by at least ~3 Ga, and that a full evaluation of progressive photosynthetic evolution against the backdrop of evolving stellar luminosity is an important topic for future work.

**Enhancement of the biological $CH_4$ cycle by a hybrid photosynthetic biosphere**

To explore the behaviour of our most primitive biosphere (Case 1), we vary the total outgassing rate of reduced gases, flux$_{volc}$, from ~0.1 to 50 Tmol $H_2$ equivalents yr$^{-1}$ while maintaining a constant atmospheric $CO_2$ (0.03 bar or ~100 times the present atmospheric level, PAL) and a constant fraction of primary producer biomass that is ultimately buried in marine sediments (2%)[28, 29] (Fig. 2). Net primary production of $H_2$-based anoxygenic phototrophs and the biogenic $CH_4$ flux to the atmosphere both predictably increase with increasing total $H_2$ outgassing flux (black solid lines with circles in Fig. 2a,b). However, an order of magnitude increase from ~1 to ~10 Tmol $H_2$ yr$^{-1}$ results in only a ~4-fold increase in the $CH_4$ flux from the biosphere. Consequently, a large input flux (>~17 Tmol $H_2$ yr$^{-1}$) is needed to maintain global average surface temperatures at or above the present value (288 K) (Fig. 2d). The requisite flux is at least ~6–10 times higher than the modern value of 1.7–2.9 Tmol $H_2$ yr$^{-1}$ proposed by Catling and Kasting[22] (neglecting the submarine $H_2S$ flux because it would readily precipitate as iron sulphide in the Archaean ferruginous ocean[22] and using a maximum estimate of modern abiotic methane flux from continental hydrothermal systems of 1.2 Tmol $H_2$ yr$^{-1}$ of ref.[22, 33]). Moreover, in the above analysis we assume a relatively high atmospheric $CO_2$ level consistent with recent estimates of 85−510 PAL at 2.77 Ga by Kanzaki and Murakami[10]. However, atmospheric $CO_2$ concentrations during the Archaean are still poorly constrained[10, 22, 34]. If previous estimates of ~10−50 PAL at 2.7 Ga[34] are correct, a larger outgassing flux is required to maintain warm (≥ 288 K) climate states (for example, ~50 Tmol $H_2$ yr$^{-1}$ for $pCO_2$ = 33 PAL; Supplementary Fig. 3). Though it is plausible that $H_2$ fluxes were higher during Archaean time[35], our results indicate that it may have been difficult to establish warm climate states with $H_2$-based photosynthesis alone.

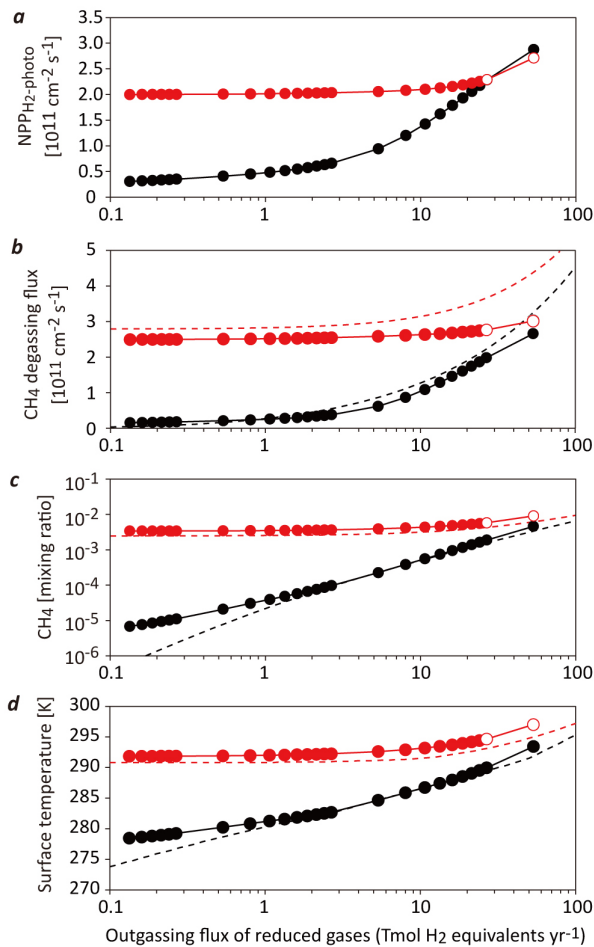

**Fig. 2. The biogeochemical response to changes in the outgassing flux of reduced gases.** Models were run with atmospheric $pCO_2$ of 100 PAL and $x$ of 2%. (**a**) Net primary production of $H_2$-using anoxygenic photosynthesis. (**b**) Biogenic $CH_4$ degassing flux from the ocean to the atmosphere. (**c**) Mixing ratio of $CH_4$ in the atmosphere. (**d**) Global surface temperature. Solid lines with circles show results obtained with a coupled atmosphere-ecosystem model, and dashed lines in (**b**), (**c**) and (**d**) represent solutions obtained by the sGRB model. Black and red lines represent results for the Case 1 and 2 biosphere, respectively. Unfilled circles indicate unconverged solutions (see Methods). Note a logarithmic scale in y-axis of (**c**).

Next, we add photoferrotrophs to the model biosphere to explore the impact of multiple coexisting forms of primitive photosynthesis (Fig. 2). We find that the addition of Fe-based photosynthesis can have a dramatic impact on the climate system. For example, at our 'default' upwelled $Fe^{2+}$ flux of ~80 Tmol Fe yr$^{-1}$ (ref.[29]), and limited Fe(III) reduction ($\gamma = 0$) average surface temperatures exceed 290 K, and are ~10–15 K warmer than the equivalent Case 1 biosphere even at reductant fluxes well above those of the modern Earth (Fig. 2d). Both the upwelled $Fe^{2+}$ flux and the degree of Fe recycling are poorly constrained, and we relax these assumptions below. Nevertheless, our results suggest that geophysical fluxes of ferrous iron may have represented an important component of Earth's early climate system, and may exert a first-order control on the climate system of reducing Earth-like planets more generally.

The marked increase in surface temperature we observed upon the inclusion of photoferrotrophy arises as a consequence of an acceleration of biogenic $CH_4$ cycling (Fig. 2c)—$CH_4$ production from photoferrotrophic organic matter promotes $H_2$-using photosynthesis and CO-consuming acetogenesis via an increase in atmospheric $CH_4$, $H_2$ and CO (Fig. 2a and Supplementary Fig. 4). Importantly, the resultant increase in biological $CH_4$ production by the Case 2 biosphere (for example, ~$2\times10^{11}$ $cm^{-2}$ $s^{-1}$ at 3 Tmol $H_2$ $yr^{-1}$; Fig. 2b), is significantly larger than the biogenic $CH_4$ flux originating directly from photoferrotrophs, (equivalent to $(1-x)/8\times$flux$_{iron}$ ~$0.37\times10^{11}$ $cm^{-2}$ $s^{-1}$), indicating a strong non-linear amplification of $CH_4$ fluxes from the biosphere upon the inclusion of photoferrotrophy. In other words, the fluxes of $CH_4$ from the biosphere that can be supported by our hybrid biosphere are larger than the additive fluxes of each form of photosynthesis acting in isolation. We suggest that this form of 'photochemical syntrophy' would have represented an important component of Earth's climate system prior to the evolution of oxygenic photosynthesis.

**Metabolic diversity promotes warm climate states**

The results of our coupled photochemistry-ecosystem model demonstrate that a hybrid ecosystem based on $H_2$ and $Fe^{2+}$ can potentially amplify global hydrogen and methane cycling and help to maintain warm surface temperatures on the early Earth. However, the magnitude of the warming effect produced by photoferrotrophs depends on several parameters that are not well constrained. In order to obtain a more comprehensive and robust understanding of the biogeochemical and radiative effects of primitive photosynthetic ecosystems, we employ a simpler parameterized model (the simplified Global Redox Balance model, hereafter 'sGRB'; see Methods), which shows the same quantitative behaviour of the coupled model (dashed lines in Fig. 2 and Supplementary Fig. 5) but has considerably reduced computational expense.

The computational efficiency of our sGRB model allows us to more fully explore conditions promoting warm climate states via a Monte Carlo (MC) analysis (see Methods). In these simulations, we vary poorly constrained parameters over a wide range of values to establish which parameter combinations allow for the maintenance of warm climate states (Table 1). We run on the order of $10^5$ simulations and only sample model runs that allow for an acceptable combination of steady-state temperature and $CH_4/CO_2$ ratio (see Methods) (Supplementary Table 4). In particular, we accept simulations that maintain warm surface temperatures ($\geq$ 288 K) and $CH_4/CO_2$ ratios below values that should result in net surface cooling due to the formation of optically thick organic hazes in the atmosphere[15, 36-41]. Although the antigreenhouse effect of hazes will depend on the optical properties of the aerosol particles[6, 36], we accept simulations with $CH_4/CO_2 \leq 0.2$ in our search for warm climates while noting that this is not meant to be a strict statement about habitability *per se* (e.g., Arney et al.[36]).

Table 1. Parameter ranges used in the Monte Carlo simulations.

| Sampled parameters | | Unit | Range | Sampling method |
|---|---|---|---|---|
| Atmospheric $CO_2$ level | $pCO_2$ | bar | $10^{-2.5}$–$10^{-1}$ | Log uniform |
| $CH_2O$ preservation efficiency | $x$ | | 0.2–20% | Log uniform |
| Outgassing flux of reduced gases | $flux_{volc}$ | Tmol $H_2$ yr$^{-1}$ | $10^{-1}$–$10^{1.5}$ | Log uniform |
| $Fe^{2+}$ upwelling flux | $flux_{iron}$ | Tmol Fe yr$^{-1}$ | $10^{0}$–$10^{3}$ | Log uniform |
| A fraction of $Fe(OH)_3$ reduced to $Fe^{2+}$ via Fe(III) reduction | $\gamma$ | | 0–$\gamma_{max}$† | Uniform |

†$\gamma_{max}$ is defined by equation (S20).

Our resampling approach allows for a robust statistical evaluation of model results, and delineates critical relationships between the photosynthetic biosphere and climate stability (Fig. 3 and Supplementary Figs. 8–10). Specifically, for our Case 1 ($H_2$-based) biosphere, the minimum $pCO_2$ consistent with a warm ($\geq$ 288 K) climate state is ~70 PAL, irrespective of large uncertainties in many model parameters. However, this also requires total $H_2$ outgassing fluxes of roughly an order of magnitude greater than the upper bound of the present value of ~3 Tmol $H_2$ yr$^{-1}$ (Fig. 3a). The acceptance rate during the MC analysis for our Case 1 biosphere is ~12% of simulations (Supplementary Table 4). If we impose a more stringent temperature requirement ($\geq$ 298 K), we find only 8 acceptable solutions after over 2,000,000 simulations for our Case 1 biosphere across the entire range of parameter space explored here. Our results thus suggest that it would have been difficult to maintain stable warm climates on the Hadean and Archaean Earth through a hydrogen-based biosphere alone. Importantly, the same is true for an entirely Fe-based photosynthetic ecosystem; if we include only photoferrotrophs as a primary producer (see Methods), the combination of parameters that can achieve warm ($\geq$ 288 K) climate states is also quite limited (Fig. 3b and Supplementary Figs. 7 and 9), and in all cases these require extremely high fluxes of $Fe^{2+}$ from the solid Earth.

However, we find that increasing metabolic diversity in the photosynthetic biosphere results in a remarkable expansion of parameter space consistent with warm climate states (Fig. 3c,d), with an increase in the acceptance rate during the MC analysis to ~24%. We also find a complex relationship between $pCO_2$ and Fe fluxes for warm climate states, such that increasing Fe fluxes can buffer climate against drops in $pCO_2$ when Fe fluxes are relatively low, but have the opposite effect when Fe fluxes are high (Fig. 3d). This relationship arises as a result of elevated $CH_4$ fluxes (and thus rising $CH_4/CO_2$ ratios) at high Fe fluxes, such that the positive radiative effects of higher atmospheric $CH_4$ are counteracted by $CH_4/CO_2$ ratios rising above the haze threshold of 0.2. Perhaps most importantly, a more diverse photosynthetic biosphere allows for stable warm climate states at much lower geophysical $Fe^{2+}$ fluxes that are likely more representative of the Archaean Earth system (Fig. 3d).

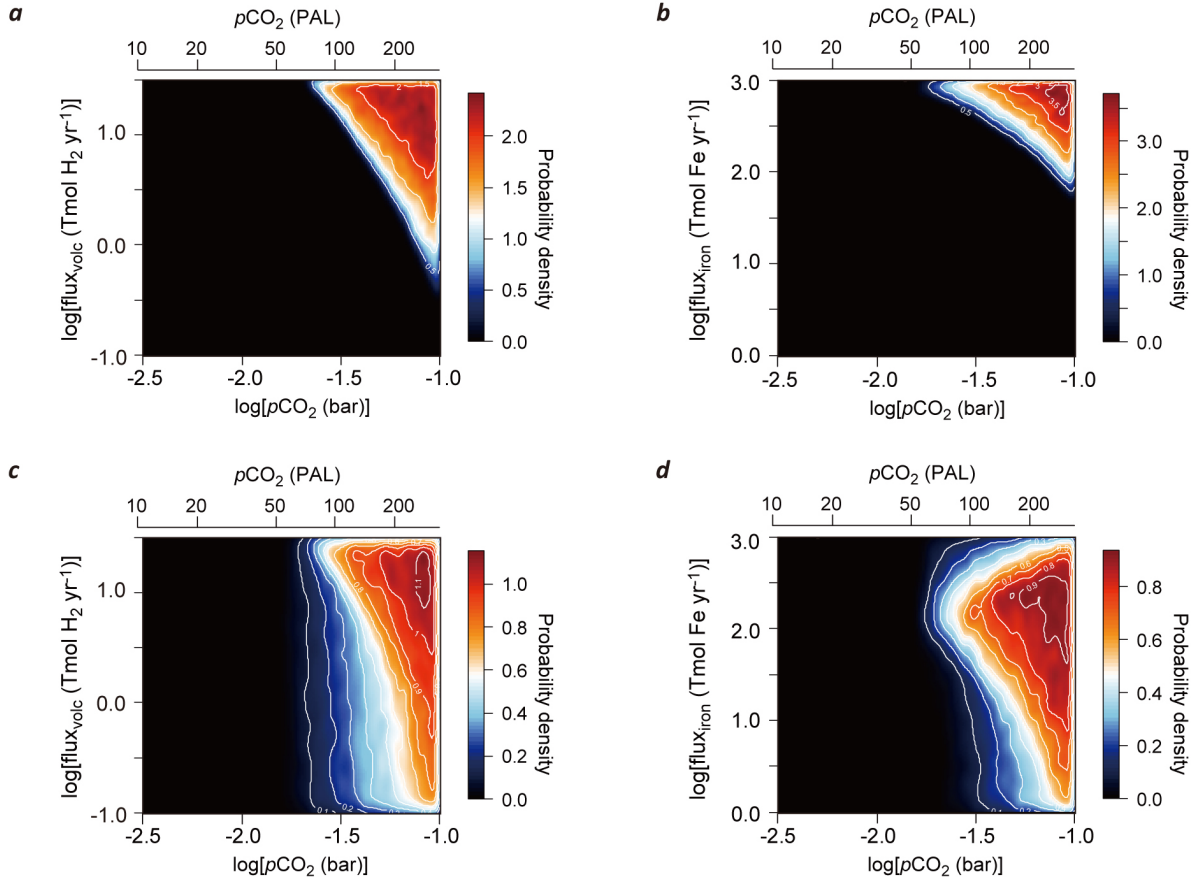

**Fig. 3. Monte Carlo simulations showing the probability density for warm (≥ 288 K) climate states with a low CH$_4$/CO$_2$ (≤ 0.2).** Probability density distribution for warm climate solutions in a phase space of total H$_2$ outgassing flux (flux$_{volc}$) and atmospheric $pCO_2$ for the Case 1 (**a**) and Case 2 biosphere (**c**). Probability density distribution in a phase space of a ferrous iron upwelling flux (flux$_{iron}$) and atmospheric $pCO_2$ for the Fe-based biosphere (see Methods) (**b**) and Case 2 biosphere (**d**). Black areas indicate parameter space where no warm solutions exist. Warm colours (orange and red) represent higher probability than cold colours (blue and white). Note a logarithmic scale in x- and y-axes.

**The role of Fe cycling in the Archaean climate system**

Our results highlight the potential importance of the geophysical and microbiological Fe cycles in regulating the climate of the early Earth and of anoxic Earth-like planets with primitive photosynthetic biospheres. In particular, for H$_2$-based or Fe-based photosynthetic biospheres the electron acceptor fluxes required to attain warm climate states at reduced luminosity are extremely high, and may require geophysical conditions that deviate beyond those of the modern or even ancient Earth – such as elevated heat flow and volcanic degassing, lower mantle oxygen fugacity, or pervasively ultramafic crustal compositions. However, our hybrid biosphere simulations yield warm climate solutions at relatively modest $Fe^{2+}$ fluxes well within those attainable on the Archaean Earth system.

For example, at the very low dissolved $SO_4^{2-}$ levels characteristic of Archaean seawater[42], typical high-temperature hydrothermal fluids would have contained dissolved $Fe^{2+}$ concentrations of ~80 mmol $kg^{-1}$ or higher, depending on hydrostatic pressure[43]. Assuming a modern axial heat flux of $3.2 \times 10^{12}$ W and an attendant water flux of $2.6 \times 10^{13}$ kg $yr^{-1}$ (ref.[44]), this yields a high temperature hydrothermal $Fe^{2+}$ flux of ~2 Tmol $yr^{-1}$. This could have been elevated by a factor of ~3–6 during Archaean time[45], and does not include any potential sources associated with off-axis alteration of ocean crust. However, results from our simulations with an Fe-based photosynthetic biosphere suggest that warm climate states would require an $Fe^{2+}$ flux on the order of ~400 Tmol $yr^{-1}$ at atmospheric $pCO_2$ values of ~100 PAL (Fig. 3b). The impact of photochemical syntrophy within the hybrid photosynthetic biosphere would thus seem to be critical for the attainment of stable warm climate states on the Archaean Earth (Fig. 3d).

In addition, in the simulations presented here we impose atmospheric $CO_2$ as a boundary condition. However, it is important to bear in mind that in reality a higher (lower) $Fe^{2+}$ flux would lead to higher (lower) atmospheric $CH_4$ levels, which should cause a reduction (increase) in atmospheric $pCO_2$ as a consequence of a climate regulation in the carbonate-silicate cycle[46]. This leads to the possibility of a novel mechanism of climate regulation involving the biogeochemical Fe cycle, as the resultant increase (reduction) in oceanic $pH$ would feed back to marine $Fe^{2+}$ concentrations by affecting the saturation states of iron species (for example, siderite). Further work utilizing an open system H-C-Fe model will be required to test this hypothesis.

In sum, our model results indicate a novel synergistic effect on early climate by primitive photosynthetic biospheres, regardless of the specific average surface temperatures required to achieve a particular climate state. We suggest that the establishment of a $H_2/Fe^{2+}$ hybrid ecosystem would have dramatically altered the mechanistic coupling between the biosphere and climate system on the early Earth. Further, the critical role of photoferrotrophs suggested by our results implies that variations in $Fe^{2+}$ availability in the ocean, rather than volcanic $H_2$ flux to the atmosphere, exerted critical control on the Archaean climate system and are likely important for other anoxic Earth-like planets. Finally, our results indicate an intriguing series of potential climate feedbacks within the H-C-Fe cycles that have not previously been explored, placing additional impetus on efforts to better understand processes regulating mass balance and internal recycling of Fe in Earth's early oceans[47-50] and the coupling between Earth's Fe and C cycles.

## Methods

Methods, including statements of data availability and any associated codes and references, are available in the online version of this paper.

**Acknowledgments:** We are grateful to J. Kasting for constructive comments on early draft of this manuscript and his sharing of FORTRAN code. This work was supported in part by JSPS grant-in-aid 25120006. K.O. acknowledges support from the NASA Postdoctoral Program, administered by the Universities Space Research Association. P.K.H. acknowledges support from TeNO/Tokyo-dome. C.T.R. acknowledges support from the NASA Astrobiology Institute and the Alfred P. Sloan Foundation.

**Author Contributions** K.O. and E.T. developed the hypothesis and designed the study. K.O. constructed the quantitative framework and performed experiments with sGRB model. P.K.H. and Y.N. carried out the experiments with the coupled model. K.O., P.K.H., and C.T.R. analysed the results. K.O., E.T. and C.T.R. wrote the paper with input from P.K.H. All authors discussed and contributed intellectually to the interpretation of the results.

**Additional information**
Supplementary Information is available in the online version of the paper. Reprints and permissions information is available online at XXX. Correspondence and requests for materials should be addressed to K.O.

**Competing financial interests**
The authors declare that no competing financial interests.

**Methods**

**Mechanistic framework.** We can construct a theoretical underpinning for the hypothetical synergetic effects of the $H_2/Fe^{2+}$ hybrid ecosystem on the $H_2$ and $CH_4$ cycles by constructing a mechanistic framework that links the biogeochemical cycles of H, C and Fe. Crucially, the analytical solution obtained below is based on the principle of redox balance in the combined ocean-atmosphere system: The sum of the source fluxes of reducing power (outgassing of reduced species from Earth's interior and burial of Fe(III) in marine sediments) should be balanced by the sink fluxes (hydrogen escape to space and burial of organic matter) at steady state (see Supplementary Methods for a full explanation). We assume that $H_2$-using photosynthesis is limited by the flux of $H_2$ gas across the air-sea interface, that $H_2$ concentrations in the surface ocean are essentially zero (due to uptake), that the activity of photoferrotrophs is limited by the upwelling flux of $Fe^{2+}$ into the surface ocean (flux$_{iron}$), that methanogenesis in the ocean interior is the main process responsible for $CH_4$ input to the atmosphere, that anaerobic oxidation of methane (AOM) by $SO_4^{2-}$ plays a minor role in the consumption of $CH_4$ in the ocean interior, and that a fraction of organic matter produced by producers, $x$, is buried in sediments. By adopting an empirical linear relationship between $CH_4$ flux to the atmosphere and $CH_4$ mixing ratio in the atmosphere (equation (S13) in Supplementary Methods)[12, 28], the following analytical solution representing primary production of organic matter via $H_2$-using photosynthesis ($NPP_{H2\text{-photo}}$, in unit of molecules cm$^{-2}$ s$^{-1}$) as a function of input fluxes of reductants in terms of $H_2$ (flux$_{volc}$) and $Fe^{2+}$ (flux$_{iron}$) to the system is obtained;

$$NPP_{H_2-photo} = \frac{1}{2} \frac{\text{flux}_{volc} + \text{flux}_{iron}/2 \cdot (1-2AB) \cdot (1-x-\gamma)}{A/v_p \alpha D + 2AB \cdot (1-x) + x} \quad (1)$$

where $\alpha$ (7.8×10$^{-4}$ M atm$^{-1}$) and $v_p$ (0.013 cm s$^{-1}$) are the Henry's law constant and a piston velocity for $H_2$, $A$ (2.5×10$^{13}$ cm$^{-2}$ s$^{-1}$), $B$ (9×10$^{-16}$ cm$^{-2}$ s$^{-1}$) and $D$ (6.02×10$^{20}$ molecules cm$^{-3}$ mol$^{-1}$ l) are constants for unit conversion, $x$ is the preservation efiency of organic matter (defined as the fraction of organic matter buried in sediments relative to primary production rate), and $\gamma$ represents a fraction of Fe(III) produced by photoferrotrophs that is reduced back to $Fe^{2+}$ via Fe(III) reduction. The coefficient of 1/2 on the right hand side represents the stoichiometric relationship between $H_2$ and CH$_2$O for H$_2$-based production ($2H_2 + CO_2 \rightarrow CH_2O + H_2O$). The second term of numerator of equation (1) represents the effect of Fe$^{2+}$-using phototrophs on H$_2$-based primary production, indicating that the activity of photoferrotrophs could raise the level of H$_2$-using photosynthesis by increasing the hydrogen deposition from the atmosphere to the ocean.

**A coupled atmosphere-ocean ecosystem model.** The photochemical model used in this study is a one-dimensional model that has been originally developed by Pavlov et al[12] and was subsequently modified by Kharecha et al.[29]. The FORTRAN code was obtained from the Virtual Planetary Laboratory (http://vpl.astro.washington.edu/sci/AntiModels/models09.html) and was modified by P.K.H and Y.N. for this study. The photochemical model includes 73 chemical species (39 long-lived species, 31 short-lived species, and 3 aerosols particles) (Supplementary Table 2) involved in 359 chemical reactions. The model includes transport by eddy and molecular diffusion to an altitude of 100 km with 1 km grid spacing. The model simulates an anoxic 1.0 bar atmosphere and variable amounts of H, C, N, S species. The solar zenith angle was set at 50°. The continuity equation was solved at each height for each of the long-lived species, including transport by eddy and molecular diffusion. The combined equations were cast in centered finite difference form. Boundary conditions for each species were applied at the top and bottom of the model atmosphere, and the resulting set of coupled differential equations was integrated to steady state using the reverse Euler method. A two-stream approach was used for the radiative transfer. The hydrogen budget, or redox budget, in the atmosphere, the ocean, and the ocean-atmosphere system is tracked during the simulation. Our simulations are run until convergence of the hydrogen budget is reached. See Supplementary Methods for further details.

**Simplified global redox balance (sGRB) model.** The analytical solution developed above (equation (1)) is useful for understanding the first-order response of the system. However, the linear fitting between CH$_4$ flux to the atmosphere (flux$_{methane}$) and atmospheric CH$_4$ mixing ratio, $f$(CH$_4$) (equation (S13)) tends to overestimate $f$(CH$_4$) when flux$_{methane}$ <1.4×10$^{11}$ cm$^{-2}$ s$^{-1}$ and to underestimate by as much as an order of magnitude when flux$_{methane}$ >3.0×10$^{11}$ cm$^{-2}$ s$^{-1}$ (Supplementary Fig. 1a), hampering quantitative discussion on the biogenic CH$_4$ cycling and climate in some cases. To develop an alternative approximation, we employed the atmospheric model of ref.[12] and performed a series of experiments with respect to the relation between flux$_{methane}$ and $f$(CH$_4$) for a wide range of flux$_{methane}$ (Supplementary Fig. 1a). We then obtained the following fit implemented in this study:

$$f(CH_4) = a \cdot \text{flux}_{methane}^{b}, \qquad (2)$$

where $a$ (=1.474×10$^{-26}$) and $b$ (=2.0291) are tunable constants. This function still tends to underestimate $f$(CH$_4$) when biogenic CH$_4$ flux is relatively high (>~2.0×10$^{11}$ cm$^{-2}$ s$^{-1}$). Nonetheless, the treatment here generally shows a good agreement with the results obtained by a coupled model of atmosphere and ocean ecosystem for a wide range of parameter settings (Fig. 2 and Supplementary Figs. 3 and 6). When we adopt equation (2) instead of equation (S13), the following relationship is obtained based on the global redox balance (see Supplementary Methods for further details):

$$\text{flux}_{volc} = \left(\frac{A}{V_p \alpha D} + x\right) \cdot \text{flux}_{marH_2} + 2Aa \cdot \left[\frac{1}{2} \cdot \left\{(1-x) \cdot \left(\frac{\text{flux}_{marH_2}}{2} + \frac{\text{flux}_{iron}}{4}\right) - \gamma \frac{\text{flux}_{iron}}{4}\right\}\right]^b - (1-x-\gamma) \cdot \frac{\text{flux}_{iron}}{2},$$

where flux$_{marH2}$ denotes the primary production of H$_2$-using photosynthesis in terms of molecules H$_2$ cm$^{-2}$ s$^{-1}$. We solve this equation with respect to flux$_{marH2}$ with a numerical scheme, giving significant savings in computational cost relative to a full coupled model while maintaining adequate quantitative accuracy. See Supplementary Discussion for further discussion of the possible errors associated this approach.

For the Case 1 biosphere, the following equation is obtained:

$$\text{flux}_{\text{volc}} = \left( \frac{A}{V_p \alpha D} + x \right) \cdot \text{flux}_{\text{marH}_2} + 2Aa \cdot \left( \frac{1-x}{4} \cdot \text{flux}_{\text{marH}_2} \right)^b.$$

We solve this equation numerically with respect to flux$_{marH2}$.

In Fig. 3b, we examine the idealized ecosystem which does not include H$_2$-using anoxygenic phototrophs. In this system H$_2$ would not be consumed by biology. Therefore the oceans would be saturated in H$_2$ and atmospheric H$_2$ level will increase until hydrogen escape can balance H$_2$ outgassing flux. The CH$_4$ outgassing rate from the ocean to the atmosphere is given by

$$\text{flux}_{\text{methane}} = \frac{1-x-\gamma}{8} \cdot \text{flux}_{\text{iron}}.$$

Given the global redox balance (equation (S15)), the mixing ratio of H$_2$ in the atmosphere is given by

$$f(\text{H}_2) = \frac{\text{flux}_{\text{volc}}}{A} - 2a \cdot \text{flux}_{\text{methane}}^b + \frac{1-x-\gamma}{2A} \text{flux}_{\text{iron}}.$$

**Monte Carlo simulation.** To assess the whole parameter uncertainty, Monte Carlo simulations were performed with the sGRB model. The uncertain variables ($p$CO$_2$, $x$, flux$_{volc}$, flux$_{iron}$, and $\gamma$) were forced within their reasonable range of values, derived from prior knowledge (Table 1, see below). Model parameters were sampled independently from uniform (non-informative) distributions, which are considered to be a reasonable starting point for objective analysis because there is not enough information for constraining the prior probability density function. By using normally distributed pseudorandom numbers in logarithm scale, except for $\gamma$, we performed a series of calculations to see whether there would be the combination of free parameters that can achieve warm climate states. Only simulations in which the calculated surface temperature and atmospheric CH$_4$/CO$_2$ met their criteria (Supplementary Table 4) were sampled as Archaean warm climate states. The sampling procedure was run for long enough to obtain the stationary distribution of probability density. For example, 423,430 MC simulations were carried out for the Case 2 biosphere until 100,000 simulations met the criterion of global surface temperature of ≥288 K with an atmospheric CH$_4$/CO$_2$ of ≤ 0.2, yielding an adoption rate of ca. 24% (Supplementary Table 4).

Given the net primary production in the modern ocean of ~45 Gt C yr$^{-1}$ (ref.[24]) and burial rate of organic carbon during the Holocene of 11.4 Tmol yr$^{-1}$, or 0.137 Gt C yr$^{-1}$ (ref.[51]), the preservation efficiency of organic matter in the near-modern ocean is estimated to be on the order of 0.3%. However, given that the pronounced cooling during the Pliocene-Pleistocene epoch has likely accelerated global erosion rate and sediment loading to the oceans by a factor of two since about 6 Myr[52], it is likely reasonable to assume that this estimate is larger than the average rate on geological timescales. On the other hand, we expect that organic matter preservation would be enhanced in the anoxic ocean interior of the Archaean Earth. The previous studies estimate the value of $x$ at 1–3% in the modern Black sea[53,54], where anoxic bottom waters enhance the preservation of organic matter. The possibility of higher values of $x$ is also discussed for the stratified anoxic, sulphate-poor environments[55]. To explore the uncertainty in the value of $x$, we set the range of $x$ at 0.2–20%. The lower bound was determined based on the fact that the model result is relatively insensitive to the value of $x$, when $x < 1\%$. The upper bound is an order of magnitude larger than the estimates for the modern Black sea. The lower bound of $\log(pCO_2)$ was set at -2.5, sufficiently low to search warm solutions. The upper bound was set at -1.0, reflecting a large uncertainty in paleosol constraint[10, 15, 34,56]. The modern total outgassing flux of reduced species was estimated to be 3.9 Tmol $H_2$ equivalents yr$^{-1}$ with a large uncertainty of at least a factor of 2 (ref.[22]). The lower values (<1 Tmol $H_2$ yr$^{-1}$) are also argued by ref.[28,57-59]. Also, there is an ongoing debate as to how the rate of volcanic outgassing has changed over time[22]. Given the large uncertainty in the total $H_2$ volcanic outgassing flux during the Archaean, we explore the value of flux$_{volc}$ in the range of $10^{-1}$–$10^{1.5}$ Tmol $H_2$ equivalents yr$^{-1}$. The global value of the ferrous iron input flux to the photic zone (flux$_{iron}$) is also highly uncertain. However, we can obtain a rough estimate as follows. The $Fe^{2+}$ concentration in the deep water was previously estimated to be in the range of ~40–120 $\mu$M[27,60]. This roughly corresponds to flux$_{iron}$ of ~70–208 Tmol Fe yr$^{-1}$, assuming the global upwelling rate of 4 m yr$^{-1}$ and oceanic area of 85% of Earth's surface area ($5.1 \times 10^8$ km$^2$). Given numerous uncertainties in these estimates, we vary the value of flux$_{iron}$ in the range of 1–1000 Tmol yr$^{-1}$. For $\gamma$, we explore its whole parameter range of 0–$\gamma_{max}$, where $\gamma_{max}$ represents the maximum value determined by flux$_{volc}$ and flux$_{iron}$ (equation (S20)).

**Climate model.** The global surface temperature was evaluated offline based on the Figure 8 of ref.[15] showing the surface temperature as a function of $pCO_2$ and $f(CH_4)$, taking into account the greenhouse effect of $C_2H_6$. We took the relationship between $pCO_2$ and surface temperature for $f(CH_4)$ of $10^{-5}$, $10^{-4}$, $10^{-3}$, $10^{-2.5}$, $10^{-2}$, and the surface temperature was evaluated as a function of $pCO_2$ and $f(CH_4)$ with a log-linear interpolation method. We consider this approach acceptable, given that our main purpose is to test the hypothesis that our 'hybrid' biosphere could help explain warm climatic condition during the Archaean by accelerating the biogenic $CH_4$ cycle, and not to completely resolve the FYSP or advocate for any particular climate state. Further work will be required to explore additional factors, such as planetary albedo, total atmospheric pressure, and the nature of aerosol particles forming the organic haze layer; none of these effects are explored in this study.

**Code availability.** The FORTRAN code of sGRB model is available from the corresponding author on reasonable request. The sGRB model is entirely reproducible using the information available in the Methods and the Supplementary Information. The code of the coupled atmosphere-ecosystem model is available upon request to the corresponding author.

**Data availability.** The data supporting the findings of this study are available within Supplementary Information files.